\documentclass[twoside,12pt]{article}

\usepackage{graphicx}
\usepackage{amssymb}
\usepackage{amsmath}

\begin{document}

\title{Quantum cosmological effects from the high redshift supernova observations
\thanks{A contribution to the International workshop on Frontiers of
Particle Astrophysics, Kiev, June 21 - 24, 2004}}
\author{V.V. Kuzmichev\\[0.5cm]
\itshape Bogolyubov Institute for Theoretical Physics,\\
\itshape National Academy of Sciences of Ukraine, \\
\itshape Kiev, 03143 Ukraine}
\date{}

\maketitle

\pagestyle{myheadings} \thispagestyle{plain} \markboth{V.V.
Kuzmichev}{Quantum cosmological effects} \setcounter{page}{1}

\begin{abstract}
Subject of this contribution is to demonstrate that the observed
faintness of the supernovae at the high redshift can be considered
as a manifestation of quantum effects at cosmological scales. We
show that observed redshift distribution of coordinate distances
to the type Ia supernovae can be explained by the local
manifestations of quantum fluctuations of the cosmological scale
factor about its average value. These fluctuations can arise in
the early universe, grow with time, and produce observed
accelerating or decelerating expansions of space subdomains
containing separate supernovae with high redshift whereas the
universe as a whole expands at a steady rate.
\end{abstract}

\newpage

\begin{center}
\Large{\textbf{Contents}}
\end{center}
\begin{enumerate}
    \item Introduction.
    \begin{itemize}
        \item Arguments in favour of the quantization of gravity
    \end{itemize}
    \item Observations
    \begin{itemize}
        \item Dark matter and dark energy problem
        \item The rate of expansion
    \end{itemize}
    \item Quantum model
    \begin{itemize}
        \item Quantization
        \item Boundary conditions and solutions
        \item The universe in the state with large quantum numbers
    \end{itemize}
    \item Link to the physical data
    \begin{itemize}
        \item Main cosmological parameters
        \item Coordinate distance to source
        \item Quantum fluctuations of scale factor
        \item Dark matter and dark energy prediction
    \end{itemize}
\end{enumerate}

\section{Introduction}
Quantum cosmology is the application of quantum theory to the
universe as a whole. Since the dominating interaction in the
cosmological realm (on the largest scales) is gravity the
extrapolation of quantum theory to the whole universe immediately
has to address the problem of quantizing the gravitational field
(e.g. \cite{Kiefer 1999}). At first glance such an attempt seems
surprising since one is used to apply quantum theory to
microscopic systems. Nevertheless one can put forward many
arguments in favour of the quantization of gravity:
\begin{itemize}
    \item \textit{Unification} (as a logical necessity and to
    avoid inconsistences). All particles are the sources of
    the gravitational field. If their gravitational fields were
    really classical, then measuring all the components of these
    fields simultaneously it would be possible to determine the
    coordinates and velocities of the particles at once, so that
    uncertainty principle will be violated. Since the
    gravitational field is coupled to all other fields, it would
    be appear strange, and even inconsistent, to have a
    drastically different framework for this one field.
    \item \textit{Singularity theorems of general relativity}.
    Under very general conditions, the occurrence of a singularity
    (where the theory breaks down) is unavoidable. Therefore a
    more fundamental theory is needed. It is expected that a
    quantum theory of gravity will be such a fundamental theory.
    \item \textit{Initial conditions in cosmology}. Since the
    singularity theorems predict the existence of an initial
    state with infinite energy density which cannot be described by
    general relativity, a quantum theory should supply a classical
    theory of gravity with appropriate initial conditions.
    \item \textit{Field-theory speculations}. It is believed that
    gravity can play the role of a regulator which can
    automatically eliminate the divergences in ordinary quantum
    field theory.
    \item \textit{Experiment-theory concordance}. A situation in
    theoretical physics is reflected in the diagram \cite{Isham 1995}:
    $$theory \ \longleftrightarrow \ concepts \
    \longleftrightarrow \ facts$$
    in which the theoretical components are linked to the physical
    data. In the case of quantum gravity it was accepted that the data which
    can be unambiguously interpreted as a result of quantum
    effects are absent, so that the diagram is shortened
    $$theory \ \longleftrightarrow \ concepts.$$
    This opinion originates from the fact that the Planck length
    $l_{Pl} = \sqrt{\frac{\hbar G}{c^{3}}}
    \simeq10^{-33}$ cm is extremely small and lies beyond the
    range of \underline{laboratory-based} \underline{experiments}.
    But it should be noticed
    that quantum effects are not a priori restricted to certain
    scales. Rather the process of decoherence \cite{Kiefer 1999} through the
    environment can explain why quantum effects are negligible or
    important for the object under consideration.

    In the present
    contribution we show that the original diagram can be restored if one
    will consider the whole \underline{universe as a laboratory} and take
    into consideration the new astrophysical data from
    supernovae type Ia observations, CMB anisotropy measurements
    (WMAP and others), HST key project, which have tremendously
    increased in volume during the last decade.

\end{itemize}

\section{Observations}
\subsection{Dark matter and dark energy problem}
Observations indicate that overwhelming majority (about 96\%) of
matter/ener\-gy in the universe is in unknown form. The observed
mass of stars gives the following values
    $$\Omega_{stars} \simeq 0.005 \quad
    \cite{Cole et al. 2001},$$
    $$\Omega_{stars} \simeq 0.003^{+0.001}_{-0.002}\quad
    \cite{Salucci Persic 1999}$$ for the density of visible
(optically bright) baryons. Observations of the cosmic microwave
background radiation (CMB) and abundances of the light elements in
the universe suggest that the total density of baryons is about
4~\% of the total energy density (e.g. \cite{Fukugita Peebles
2004})
    $$\Omega_{B} \simeq 0.04,\qquad \qquad
    \Omega_{B}/\Omega_{stars} \sim O(10).$$
This value is one order greater than the observed mass of stars.
It means that most of baryonic matter today is not contained in
stars and is invisible (dark).

The CMB anisotropy measurements allow to determine the total
energy density $\Omega_{tot}$ and the matter component
$\Omega_{M}$. The recent data give the strong evidence that the
present-day universe is spatially flat (or very close to it):
    $$\Omega_{tot} \simeq 1$$
\cite{de Bernardis et al. 2000}, \cite{Netterfield et al. 2002},
\cite{Pryke et al. 2002}, \\ \cite{Sievers et al. 2003},
\cite{Spergel et al. 2003} and the mean matter density equals
about 30~\% of the total energy density. The independent
information about the mean matter density $\Omega_{M}$ extracted
from the high redshift supernovae Ia data on the assumption that
the universe is spatially flat gives the close values:
    $$\Omega_{M} = 0.29 ^{+0.05}_{-0.03}\quad
    \cite{Riess et al. 2004}.$$
The discrepancies between the matter density $\Omega_{M}$ and the
density of baryons $\Omega_{B}$ on the one hand and the total
energy density $\Omega_{tot}$ and the matter density $\Omega_{M}$
on the other hand are signs that there must exist non-baryonic
dark matter with the density
    $$\Omega_{DM} \simeq 0.25,$$
and some mysterious cosmic substance (so-called dark energy) with
the density
    $$\Omega_{X} \simeq 0.71.$$
The origin and composition of both dark matter and dark energy are
unknown.

Dark matter manifests itself in the universe through the
gravitational interaction. Its presence allows to explain rotation
curves for galaxies and large-scale structure of the universe in
the models with standard assumption of adiabatic density
perturbations. As regards dark energy it is worth mentioning that
its expected properties are unusual. It is unobservable (in no way
could it be detected in galaxies) and spatially homogeneous.

\subsection{The rate of expansion}
The observed faintness of the type Ia supernovae (SNe) at the high
redshift attracts cosmologists' attention in connection with the
hypothesis of an accelerating expansion of the present-day
universe proposed for its explanation. Such a conclusion assumes
that dimming of the supernovae is hardly caused by physical
phenomena non-related to overall expansion of the universe as a
whole, such as unexpected luminosity evolution, effects of
contaminant gray intergalactic dust, gravitational lensing, and
others (e.g. \cite{Tonry et al. 2003}).

Furthermore it is supposed that matter component of energy density
in the universe $\rho_{M}$, which includes visible and invisible
(dark) baryons and dark matter, varies with the expansion of the
universe as $a^{-3}$ (that is it has practically vanishing
pressure), where $a$ is a cosmological scale factor,
    $$\rho_{M} \sim a^{-3} \quad (p_{M} \approx 0),$$
while dark energy is described by the following equation of state
    $$p_{X} = w_{X} \rho_{X},\quad \mbox{where}\ -1 \leq w_{X} \leq -
    \frac{1}{3}.$$
Parameter $w_{X}$ can be constant, as e.g. in the models with the
cosmological constant ($\Lambda$CDM-models), or may vary with time
as in the rolling scalar field scenario (models with
quintessence).

Even if regarding baryon component one can assume that the
pressure of baryons may be neglected due to their relative small
amount in the universe, for dark matter (whose nature and
properties can be extracted only from its gravitational action on
ordinary matter) such a dependence on the scale factor may not
hold in the universe taken as a whole (in contrast to local
manifestations, for example in large-scale structure formation).
Since the contribution from all baryons into the total energy
density does not exceed 4~\% the evolution of the universe as a
whole is determined mainly by dark matter and dark energy.

Subject of this contribution is to demonstrate that the observed
dimming of the supernovae at the high redshift can be considered
as a manifestation of \textit{quantum effects} at cosmological
scales.

\section{Quantum model}
\subsection{Quantization}
Just as in ordinary quantum nonrelativistic and relativistic
theories one can assume that the problem of evolution and
properties of the universe as a whole in quantum cosmology should
be reduced to the solution of the functional partial differential
equation determining the eigenvalues and the eigenstates of some
hamiltonian-like operator (in space of generalized variables,
whose roles are played by the metric tensor components and matter
fields).

For simplicity we restrict our study to the case of minimal
coupling between geometry and the matter. Considering that scalar
fields play a fundamental role both in quantum field theory and in
the cosmology of the early universe we assume that, originally,
the universe is filled with matter in the form of a scalar field
$\phi$ with some potential energy density  $V(\phi)$.

Let us consider homogeneous and isotropic universe with positive
spatial curvature. Assuming that the scalar field $\phi$ is
uniform and the geometry is defined by the Robertson-Walker
metric, we represent the action functional in the conventional
form
\begin{eqnarray}
  S  = \int d\eta \,\left[\pi _{a}\,\partial_{\eta}a + \pi _{\phi }\,
  \partial_{\eta}\phi  -  \mbox{H} \right].
\label{1}
\end{eqnarray}
Here $\eta$ is the time parameter (that is related to the
synchronous proper time $t$ by the differential equation $dt =
N\,a\,d\eta$), $a(\eta)$ is a scale factor; $\pi _{a}$ and $\pi
_{\phi}$ are the momenta canonically conjugate with the variables
$a$ and $\phi$, respectively. The Hamiltonian $\mbox{H}$ is
following
\begin{equation}
    \mbox{H} = \frac{1}{2}\,N\,\left[ -\,\pi _{a}^{2}
       + \frac{2}{a^{2}}\,\pi _{\phi }^{2} - a^{2}
       + a^{4}\,V(\phi ) \right] \equiv N\, {\cal R},
\label{2}
\end{equation}
where $N(\eta)$ is a function that specifies the time-reference
scale. Here and below we use the modified Planck units:
    $$l_{P} = \sqrt{2G\hbar/(3\pi c^{3})} = 0.74 \times 10^{-33}\ \mbox{cm},$$
    $$\rho_{P} = 3c^{4}/(8 \pi G l_{P}^{2}) = 1.63 \times 10^{117}\ \mbox{GeV cm}^{-3}.$$
The function $N$ plays the role of a Lagrange multiplier, and the
variation with respect to $N$ leads to the constraint equation
    $$\delta S/\delta N = 0 \quad \Rightarrow \quad {\cal R} = 0.$$
The structure of the constraint is such that true dynamical
degrees of freedom cannot be singled out explicitly. In the model
being considered, this difficulty is reflected in that the choice
of the time variable is ambiguous (so called problem of time). For
the choice of the time coordinate to be unambiguous, the model
must be supplemented with a coordinate condition. When the
coordinate condition is added to the field equations, their
solution can be found for chosen time variable. However, this
method of removing ambiguities in specifying the time variable
does not solve the problem of a quantum description.

Therefore we shall use another approach and remove the above
ambiguity  with the aid of a coordinate condition imposed prior to
varying the action functional. The invariance of action is
restored by parametrizing the action. This approach formally
agrees with the procedure of transformation from the
Wheeler-DeWitt equation to a functional Schr\"{o}dinger equation
(from the Arnowitt, Deser and Misner to the Kucha\v{r}
description) widely discussed in literature (e.g. \cite{Kuchar
Torre 1991}, \cite{Ambrus Hajicek 2001}).

We will choose the coordinate condition in the form
\cite{Kuzmichev 1998}, \cite{Kuzmichev 1999}, \cite{Kuzmichev
Kuzmichev 2002}
\begin{equation}
  g^{00}\left(\partial_{\eta}T\right)^{2} = \frac{1}{a^{2}}\,, \quad
  \mbox{or} \quad \partial_{\eta}T = N,
  \label{3}
\end{equation}
where $T$ is the privileged time coordinate, and include it in the
action functional with the aid of a Lagrange multiplier $P$
\begin{equation}
  S = \int \! d\eta\, \left[\,\pi _{a}\,\partial_{\eta}a + \pi _{\phi }\,
  \partial_{\eta}\phi + P\,\partial_{\eta}T - {\cal H}\,\right],
\label{4}
\end{equation}
where
\begin{equation}
 {\cal H} = N\,[\,P + {\cal R}\,]
\label{5}
\end{equation}
is the new Hamiltonian. The constraint equation reduces to the
form
\begin{equation}
  P + {\cal R} = 0.
  \label{6}
\end{equation}
Parameter $T$ can be used as an independent variable for the
description of the evolution of the universe.

In quantum theory, the constraint equation comes to be a
constraint on the wave function that describes the universe filled
with a scalar field and radiation. The time-dependent equation has
a following form
\begin{equation}\label{7}
    i\,\partial_{T} \Psi = \hat{\mathcal{H}} \Psi,
\end{equation}
with a Hamiltonian-like operator
\begin{equation}\label{8}
    \hat{\mathcal{H}} = \frac{1}{2} \left(\partial_{a}^{2} -
    \frac{2}{a^{2}}\,\partial_{\phi}^{2} - a^{2} + a^{4} V(\phi)
    \right ).
\end{equation}
The wavefunction $\Psi$ depends on the cosmological scale factor
$a$, scalar field $\phi$, and time coordinate $T$. One can
introduce, at least formally, a positive definite scalar product
$\langle \Psi | \Psi \rangle < \infty $ and specify the norm of a
state. This makes it possible to define a Hilbert space of
physical states and to construct quantum mechanics for model of
the universe being considered.

Eq.~(\ref{7}) allows a particular solution with separable
variables
\begin{equation}\label{9}
    \Psi = \mbox{e}^{\frac{i}{2} E T} \psi_{E},
\end{equation}
where the function $\psi_{E}$ is given in $(a,\phi)$-space of two
variables and satisfies the time-independent equation
\begin{equation}\label{10}
 \left( -\,\partial _{a}^{2} +  a^{2} - a^{4} \hat{\rho}_{\phi}  - E  \right)
 \psi _{E} = 0.
\end{equation}
Here the operator
\begin{equation}\label{11}
\hat{\rho}_{\phi} = -\, \frac{2}{a^{6}}\,\partial _{\phi }^{2} +
V(\phi)
\end{equation}
corresponds to the energy density of the scalar field in classical
theory. The eigenvalue $E$ determines the components of the
energy-momentum tensor
\begin{equation}\label{12}
  \widetilde T^{0}_{0} = \frac{E}{a^{4}},\quad
 \widetilde T^{1}_{1} = \widetilde T^{2}_{2} = \widetilde T^{3}_{3} =
 -\,\frac{E}{3\, a^{4}},\quad \widetilde T^{\mu }_{\nu } = 0
  \ \ \mbox{for} \ \  \mu \neq \nu.
\end{equation}
We shall consider the case $E > 0$ and call a source determined by
the energy-momentum tensor $\widetilde T^{\mu}_{\nu}$ a radiation.

Eq.~(\ref{10}) turns into the Wheeler-DeWitt equation for the
minisuperspace model in the special case $E = 0$.

\subsection{Boundary conditions and solutions}
A solution to equation (\ref{10}) can  be represented as a
superposition of the functions $\varphi _{\epsilon }$ of the
adiabatic approximation \cite{Kuzmichev Kuzmichev 2002}
\begin{eqnarray}\label{13}
    \psi _{E}(a, \phi ) = \int_{- \infty}^{\infty }\! d\epsilon \,
    \varphi _{\epsilon }(a, \phi )\, f_{\epsilon}(\phi ; E),
\end{eqnarray}
with
\begin{equation}\label{14}
  \left(-\, \partial _{a}^{2} + U \right) \varphi _{\epsilon } =
  \epsilon \,\varphi _{\epsilon }.
\end{equation}
Here
\begin{equation}\label{15}
U = a^{2} - a^{4} V(\phi)
\end{equation}
is the effective potential with the turning points $a_{i} =
a_{i}(\epsilon, \phi)$: $U(a_{i}) = \epsilon, \ a_{1} < a_{2}$.

In order to specify the solution of Eq.~(\ref{14}) at given
potential of the scalar field $V(\phi)$, it has to be supplemented
by boundary conditions.

The effective potential $U$ as a function of the scale factor $a$
has a form of the barrier. Therefore the general solution of
Eq.~(\ref{14}) outside the barrier can be represented in the form
of the superposition of the wave incident upon the barrier,
$\varphi_{\epsilon}^{(-)}(a)$, and the outgoing wave,
$\varphi_{\epsilon}^{(+)}(a)$. We have
\begin{equation}\label{16}
  \varphi _{\epsilon }(a) = {\cal A}(\epsilon )\,\varphi _{\epsilon
  }^{(0)}(a) \quad \mbox{for} \ 0 < a < {\cal R},
\end{equation}
\begin{eqnarray}\label{17}
 \varphi_{\epsilon}(a)  =
 \frac{1}{\sqrt{2\pi}}\,\left[\varphi_{\epsilon}^{(-)}(a)
  - {\cal S}(\epsilon)\,\varphi_{\epsilon}^{(+)}(a) \right] \quad
 \mbox{for} \ a > {\cal R} > a_{3}.
\end{eqnarray}
where $U(a_{3}) = 0$, $a_{3} > a_{2}$.

The function  $\varphi _{\epsilon}^{(0)}(a)$ is the solution of
Eq.~(\ref{14}) that is regular at the origin, $a = 0$, and weakly
dependent on $\epsilon$. It can be normalized to unity.

Beyond the turning points in semiclassical (WKB) approximation the
incident and outgoing waves can be written in an explicit form
\begin{eqnarray}\label{18}
  \varphi_{\epsilon}^{(\pm)}(a) = \frac{1}{\sqrt{2}\,(\epsilon -
  U)^{1/4}} \exp \left\{\mp\,i \int_{a_{2}}^{a}\! \sqrt{\epsilon -
  U}\,da \pm \frac{i\pi}{4}\right\}.
\end{eqnarray}

The amplitude ${\cal A}(\epsilon )$ of the wavefunction inside the
barrier and the amplitude ${\cal S}(\epsilon)$ (an analog of
S-matrix) show a resonance behaviour, that is they have a sharp
peak at $\epsilon = \epsilon_{n}$, while the resonance curve has a
width $\Gamma_{n}$,
\begin{equation}\label{19}
 \left|{\cal A}(\epsilon )\right|^{2} \simeq \frac{1}{\pi}
 \frac{\Gamma_{n}}{(\epsilon - \epsilon_{n})^{2} +
 \Gamma_{n}^{2}},
\end{equation}
\begin{equation}\label{20}
 {\cal S}(\epsilon) = \exp [2i\delta(\epsilon)],\quad
 \delta(\epsilon) = \sigma(\epsilon) + \delta_{res}(\epsilon),
\end{equation}
\begin{equation}\label{21}
 \sigma(\epsilon) = \frac{1}{2i} \ln
 \frac{\varphi_{\epsilon}^{(-)}({\cal R})}{\varphi_{\epsilon}^{(+)}({\cal
 R})},\quad
 \delta_{res}(\epsilon) = \arctan \frac{\Gamma_{n}}{\epsilon -
 \epsilon_{n}}.
\end{equation}
Using the explicit forms of the incident and outgoing waves we
find
    $$\sigma(\epsilon) = \int_{a_{2}}^{{\cal R}}\!\! \sqrt{\epsilon - U}\, da - \frac{\pi}{4}.$$
The width is equal to
\begin{equation}\label{22}
  \Gamma_{n} \simeq \exp \left\{-2 \int_{a'_{1}}^{a'_{2}}\!\!
  \sqrt{U - \epsilon_{n}}\, da \right\}\quad \mbox{at} \quad
  \Gamma_{n} \ll \epsilon_{n},
\end{equation}
where $a'_{i} = a_{i}(\epsilon_{n})$.

The parameters $\epsilon_{n} > 0$ (position of the level) and
$\Gamma_{n} > 0$ (its width), $n = 0, 1, 2 \ldots $ (number of the
state) describe the universe in $n$-th quasistationary state. In a
wide variety of quantum states of the universe, described by the
time-independent equation (\ref{10}), quasistationary states are
the most interesting, since the universe in such states can be
characterized by the set of standard cosmological parameters. At
small width, $\Gamma_{n} \ll 1$, the wavefunction of the
quasistationary state $\varphi_{\epsilon}(a)$ as a function of $a$
has a sharp peak for $\epsilon = \epsilon_{n}$ and it is
concentrated mainly in the region limited by the barrier $U$,
\begin{equation}\label{23}
  \left|\varphi_{\epsilon_{n}}\right|_{a < {\cal R}} \sim
  \left(\frac{2}{{\cal R}\,
  \Gamma_{n}}\right)^{1/2}\,\left|\varphi_{\epsilon_{n}}\right|_{a > {\cal
  R}}.
\end{equation}

If $\epsilon \neq \epsilon_{n}$, then at small width, $\Gamma_{n}
\ll 1$, the wavefunction reaches the great values on the boundary
of the barrier, while under the barrier it is small,
$\varphi_{\epsilon} \sim O(\Gamma_{n})$,
\begin{equation}\label{24}
  \left|\varphi_{\epsilon}\right|_{max}^{2} \sim
  \frac{\Gamma_{n}}{{\cal R}}\,\frac{\sqrt{\epsilon}}{(\epsilon -
  \epsilon_{n})^{2}}\, \left|\varphi_{\epsilon}\right|_{a =
  a_{3}}^{2}.
\end{equation}

Therefore following \cite{Fock 1976} one can introduce some
approximate function which is equal to exact wavefunction inside
the barrier and vanishes outside it. This function can be
normalized and used in calculations of expectation values. Such an
approximation does not take into account exponentially small
probability of tunneling through the barrier $U$. It is valid for
calculation of observed parameters within the lifetime of the
universe, when the quasistationary states can be considered as
stationary ones with $\epsilon = \epsilon_{n}$.

\subsection{The universe in the state with large quantum\\ numbers}
We shall assume that the average value of the scale factor
$\langle a \rangle$ in the state with large quantum numbers
determines the scale factor of the universe in classical
approximation. Then the time-independent equation (\ref{10}) can
be reduced to the form of the first Einstein-Friedmann equation in
terms of average values (for details see \cite{Kuzmichev Kuzmichev
2004a})
\begin{equation}\label{25}
    \left(\frac{1}{\langle a \rangle}\,\frac{d \langle a
    \rangle}{dt}\right)^{2} = \langle \rho_{tot} \rangle - \frac{1}{\langle a
    \rangle^{2}},
\end{equation}
where
\begin{equation}\label{26}
    \langle \rho_{tot} \rangle =
    \frac{2}{\langle a \rangle ^{6}} \left \langle -\,\partial_{\phi}^{2} \right
    \rangle + \left \langle V \right \rangle + \frac{E}{\langle a \rangle ^{4}}
\end{equation}
is the mean total energy density.

The quantum state of the universe depends on the form and the
value of the potential $V(\phi)$. Just as in classical cosmology
which uses a model of the slow-roll scalar field in quantum theory
based on the time-independent equation (\ref{10}) it makes sense
to consider a scalar field $\phi$ which slowly evolves (in
comparison with a large increase of the average value of the scale
factor $\langle a \rangle$) into a vacuum-like state with zero
energy density, $V(\phi_{vac}) = 0$, from some initial state
$\phi_{start}$ with Planck energy density, $V(\phi_{start}) \sim
\rho_{P}$. The latter condition allows us to consider the
evolution of the universe in time in classical sense. Reaching the
vacuum-like state $\phi_{vac}$ the scalar field $\phi$ begins to
oscillate about the equilibrium vacuum value due to the quantum
fluctuations. Here the potential $V(\phi)$ of the scalar field can
be well approximated by the potential of harmonic oscillator
\cite{Kuzmichev Kuzmichev 2004b}
    $$V(\phi) = \frac{m^{2}}{2}\left(\phi - \phi_{vac}\right )^{2}
    + \ldots,$$
where $m^{2} = \left ( d^{2} V/d \phi^{2} \right )_{\phi_{vac}} >
0$. The oscillations in such a potential well can be quantized.
The spectrum of energy states of the scalar field $\phi$ obtained
here has the following form: $M = m \left (s + \frac{1}{2} \right
)$, where $m$ is a mass of elementary quantum excitation of the
vibrations of the scalar field, while $s$ counts the number of
these excitations. The value $M$ can be treated as a quantity of
matter/energy in the universe.

In the states of the universe with large quantum numbers, $n \gg
1$ and $s \gg 1$, we have the following relations
\begin{equation}\label{28}
    E = 4 \langle a \rangle \left [\langle a \rangle - M
    \right ],
\end{equation}
\begin{equation}\label{29}
    \langle \rho_{tot} \rangle = \gamma \, \frac{M}{\langle a \rangle^{3}} +
    \frac{E}{\langle a \rangle^{4}},
\end{equation}
where the coefficient $\gamma = 193/12$ arises in calculation of
expectation value for the operator of energy density of scalar
field and takes into account its kinetic and potential terms.

\section{Link to the physical data}
\subsection{Main cosmological parameters}
In matter dominated universe $M \gg E/(4 \langle a \rangle)$ and
the quantity of matter/ener\-gy $M$ and the mean energy density
$\langle \rho_{tot} \rangle$ in the universe taken as a whole
(that is in quantum states which describe only homogenized
properties of the universe) satisfy the following relations
\begin{equation}\label{30}
    M = \langle a \rangle, \qquad
    \langle \rho_{tot} \rangle = \frac{\gamma }{\langle a
    \rangle^{2}}.
\end{equation}
It is interesting to examine these relations for the parameters of
the present-day universe (the mean energy density $\rho_{0}$, the
mass of the observed part of the universe $M_{0}$, radius of
curvature or distance to the particle horizon $a_{0}$, the age of
the universe $t_{0}$)
    $$\rho_{0} \sim 10^{-29}\ \mbox{g cm}^{-3}, \quad M_{0} \sim 10^{80}\ \mbox{GeV},
    \quad a_{0} \sim 10^{28}\ \mbox{cm}, \quad t_{0} \sim 10^{17}\
    \mbox{s}.$$
In modified Planck units we have
\begin{equation}\label{31}
    M_{0} \sim a_{0} \sim t_{0} \sim 10^{61},
\end{equation}
while the total energy density will be the following
\begin{equation}\label{32}
    \rho_{0} \sim \frac{1}{a_{0}^{2}} \sim \frac{1}{t_{0}^{2}} \sim
    10^{-122}.
\end{equation}
A good agreement between the theory and the observations should be
pointed out at once.

Our quantum model predicts that the dimensionless age parameter is
equal to unity
\begin{equation}\label{33}
    \langle a \rangle \sim t,\quad H t = 1.
\end{equation}
This agrees with the observations:
    $$0.72 \lesssim H_{0} t_{0} \lesssim 1.17 \quad \cite{Peebles Ratra 2003},$$
    $$H_{0} t_{0} = 0.96 \pm 0.04 \quad \cite{Tonry et al. 2003},$$
    $$H_{0} t_{0} \simeq 0.93 \quad \cite{Krauss 2003},$$
    $$H_{0} t_{0} \simeq 0.995 \quad \cite{Spergel et al. 2003}.$$
The quantum theory also predicts that the universe in highly
excited states is spatially flat to within about 7~\%,
    $$\Omega_{tot} = 1.066.$$
It is in harmony with the latest observations as well:
    $$\Omega_{tot} = 1 \pm 0.12 \quad \cite{de Bernardis et al. 2000},$$
    $$\Omega_{tot} = 1.02 \pm 0.06 \quad \cite{Netterfield et al. 2002},$$
    $$\Omega_{tot} = 1.04 \pm 0.06 \quad \cite{Pryke et al. 2002},$$
    $$\Omega_{tot} = 0.99 \pm 0.12 \quad \cite{Sievers et al. 2003},$$
    $$\Omega_{tot} = 1.02 \pm 0.02 \quad \cite{Spergel et al. 2003}.$$

\subsection{Coordinate distance to source}
Let us consider the problem of observed faintness of type Ia
supernovae at the high redshift within the framework of our
quantum approach.

A luminosity distance $d_{L}$ is connected with the distance to
source in comoving reference frame $r(z)$ by a following simple
relation,
    $$d_{L} = (1 + z)\,r(z), \qquad
    f_{obs} = \frac{L}{4 \pi d_{L}^{2}},$$
where $f_{obs}$ is the measured flux, $L$ is the luminosity of the
standard candle, $z \equiv \Delta \lambda / \lambda =
a_{0}/\langle a \rangle - 1$ is the cosmological redshift. The
distance to source in comoving reference frame is determined via
the expansion rate $H(z)$
\begin{eqnarray}
    \nonumber
  r(z) &=& a_{0} \, \sin \left (\frac{1}{a_{0}} \int_{0}^{z}\!\! \frac{dz}{H(z)}\right )
  \quad \mbox{for} \quad \Omega_{tot} > 1,\\
    \label{34}
  r(z) &=& \int_{0}^{z}\!\! \frac{dz}{H(z)} \quad \mbox{for} \quad \Omega_{tot} = 1,\\
    \nonumber
  r(z) &=& a_{0} \, \sinh \left (\frac{1}{a_{0}} \int_{0}^{z}\!\! \frac{dz}{H(z)}\right )
  \quad \mbox{for} \quad \Omega_{tot} < 1.
\end{eqnarray}
In our quantum model in the case of a flat universe the
dimensionless coordinate distance obeys the logarithmic law
\cite{Kuzmichev Kuzmichev 2004a}
\begin{equation}\label{35}
     H_{0}\,r(z) = \ln (1 + z).
\end{equation}
In Figure 1 the dimensionless coordinate distance $H_{0} r(z)$ as
a function of redshift $z$ is shown. Our quantum model is drawn as
a lower red line. It describes the expansion of the universe as a
whole at a steady rate (with vanishing deceleration parameter
$q(z) \equiv - \ddot{a} a/\dot{a}^{2}$). The upper blue line
corresponds to the model with dark energy in the form of
cosmological constant whose contribution to the total energy
density is equal to 70~\%. The latter phenomenological model
predicts an accelerated expansion of the present-day universe with
the following deceleration parameter: $q_{0} = - 0.55$.

\begin{figure}[ht]
\begin{center}
\includegraphics*[scale=1.4]{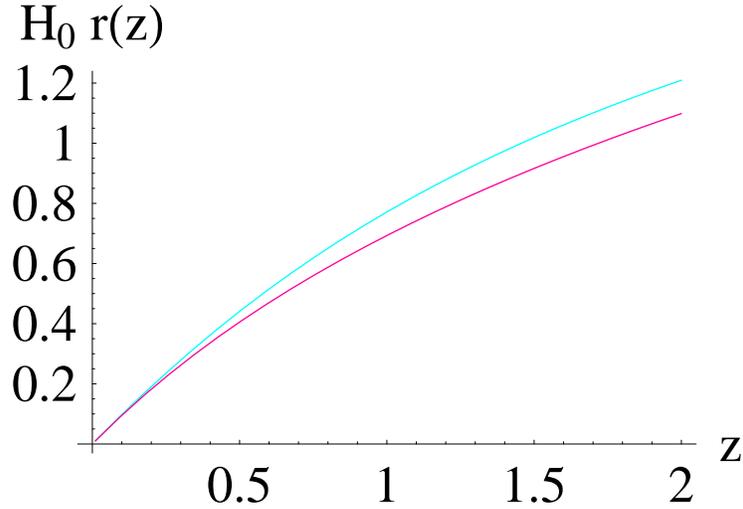}
\end{center}
\caption{Dimensionless coordinate distance $H_{0}\, r(z)$ vs. $z$
in quantum model (lower red line) and in $\Lambda$CDM-model with
$\Omega_{X} = 0.7$ and present-day deceleration parameter $q_{0} =
- 0.55$ (upper blue line).} \label{fig:1}
\end{figure}

In Figure 2 extra middle green line is added. It corresponds to
the model with cosmological constant whose contribution to the
total energy density is equal to 56~\% . The present-day
deceleration parameter for such a model is the following: $q_{0} =
- 0.34$. The rest as in Figure 1. The red line which represents
our quantum model practically coincides with the green line of the
model with smaller acceleration. But it is worth to note that
these two models describe the different physics.

\begin{figure}[ht]
\begin{center}
\includegraphics*[scale=1.4]{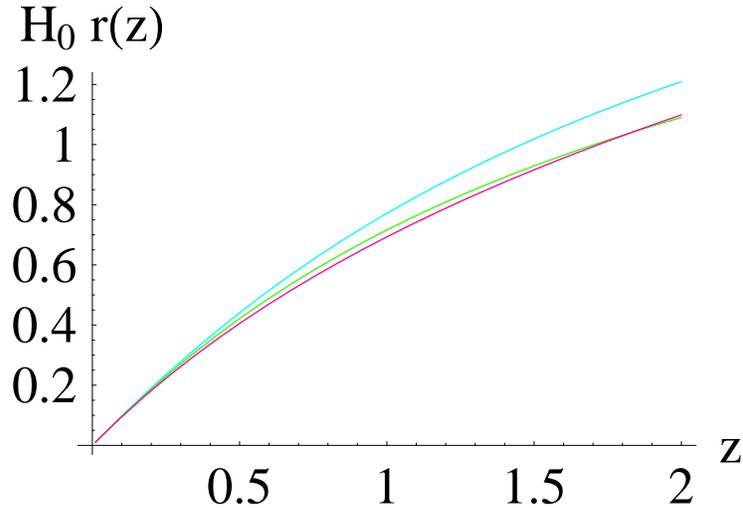}
\end{center}
\caption{Dimensionless coordinate distance $H_{0}\, r(z)$ vs. $z$
in quantum model (lower red line) and in $\Lambda$CDM-models with
$\Omega_{X} = 0.56$ and $q_{0} = - 0.34$ (middle green line) and
$\Omega_{X} = 0.7$ and $q_{0} = - 0.55$ (upper blue line).}
\label{fig:2}
\end{figure}

In Figure 3 the three above mentioned models are compared with the
observational data. The type Ia supernovae are shown as solid
circles. From more then 170 objects of the survey we have drawn
only a few dozens of typical supernovae which allow to follow the
general tendency.

\begin{figure}[ht]
\begin{center}
\includegraphics*[scale=1.4]{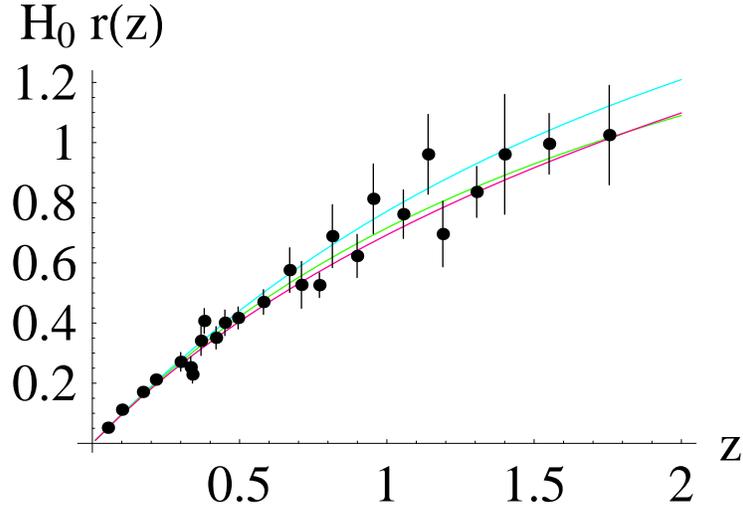}
\end{center}
\caption{Dimensionless coordinate distances $H_{0}\, r(z)$ to
supernovae at redshift $z$. The observed SNe Ia \cite{Riess et al.
2004}, \cite{Daly Djorgovski 2004} are shown as solid circles. The
rest as in Fig.~2.} \label{fig:3}
\end{figure}

We may conclude that our quantum model is entirely consistent with
the data of observations. This conclusion agrees with the result
of data processing by \cite{Daly Djorgovski 2004} who demonstrate
that the model with lower value of the contribution from dark
energy in the form of cosmological constant (the green line in
Figures 2 and 3) may be preferred.

\subsection{Quantum fluctuations of scale factor}
Deviations of the coordinate distances $H_{0}\,r(z)$ from the
logarithmic law (\ref{35}) towards both larger and smaller
distances for some supernovae can be explained by the local
manifestations of quantum fluctuations of scale factor about its
average value $\langle a \rangle$. Such fluctuations arose in the
Planck epoch ($t \sim 1$) due to finite widths of quasistationary
states. They can cause the formation of nonhomogeneities of matter
density which have grown with time into the observed large-scale
structures in the form superclusters and clusters of galaxies,
galaxies themselves etc. \cite{Kuzmichev Kuzmichev 2002}.

Let us consider the influence of mentioned fluctuations on visible
positions of supernovae. The position of quasistationary state can
be determined only approximately, $|\delta \epsilon_{n}| \sim
\Gamma_{n}$, and the scale factor of the universe in the $n$-th
state can be found only with an uncertainty $\delta a \gtrless 0$,
    $$\epsilon_{n} \rightarrow \epsilon_{n} + \delta \epsilon_{n}
    \qquad  \Rightarrow
    \qquad \langle a \rangle \rightarrow \langle a \rangle + \delta
    a.$$
If one assumes that just the fluctuations of the scale factor
$\delta a$ cause deviations of positions of sources at high
redshift from the logarithmic law (\ref{35}), then the coordinate
distances will be given by the following expression
\begin{equation}\label{36}
    H_{0}\,r(z) = \ln \left[\left(1 +
    \frac{\delta a}{\langle a \rangle} \right)^{-1} (1 + z)\right].
\end{equation}
The possible values of coordinate distances in quantum model which
takes into account fluctuations are shown as an area between two
orange lines in Figure 4. Practically all supernovae fall within
the shown limits.

\begin{figure}
\begin{center}
\includegraphics*[scale=1.4]{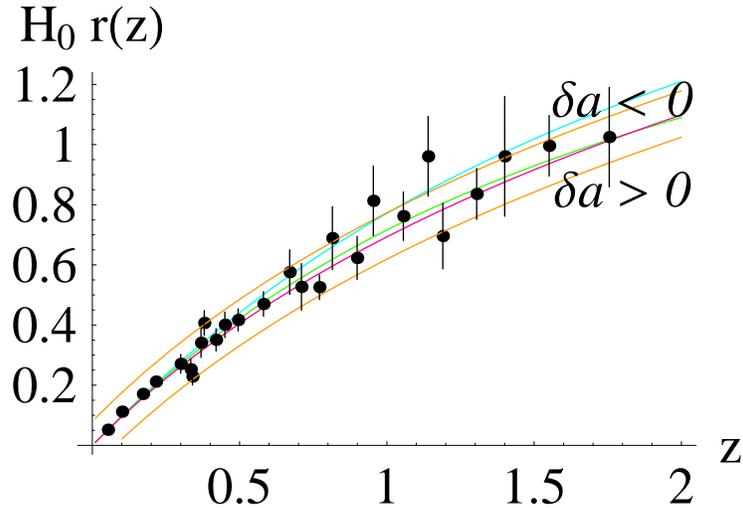}
\end{center}
\caption{Dimensionless coordinate distances $H_{0}\, r(z)$ to
supernovae at redshift $z$. An area between two orange lines
corresponds to possible values of coordinate distances in quantum
model with fluctuations.} \label{fig:4}
\end{figure}

Thus the observed faintness of the SNe Ia can in principle be
explained by the logarithmic-law dependence of coordinate distance
on redshift in generalized form (\ref{36}) which takes into
account the fluctuations of scale factor about its average value.
These fluctuations can arise in the early universe and grow with
time into observed deviations of the coordinate distances of
separate supernovae at the high redshift. They produce
accelerating or decelerating expansions of space subdomains
containing such sources whereas the universe as a whole expands at
a steady rate.

The same analysis one can make for radio galaxies as well.

One can come to a conclusion that the universe as a whole expands
at a steady rate analyzing the information about the expansion
rate extracted directly from the data on coordinate distances to
supernovae and radio galaxies.

In Figure 5 the derived values of the dimensionless expansion rate
$E(z) \equiv (\dot{a}/a) H_{0}^{-1}$ obtained by \cite{Daly
Djorgovski 2004} are shown. The inaccuracy of measurements and
uncertainties in data processing do not allow to take this plot as
a final outcome, but only look at the global trends. These trends
are supported by the results of our calculations drawn in Figure 5
as an area between the red lines.

\begin{figure}[ht]
\begin{center}
\includegraphics*[scale=0.7]{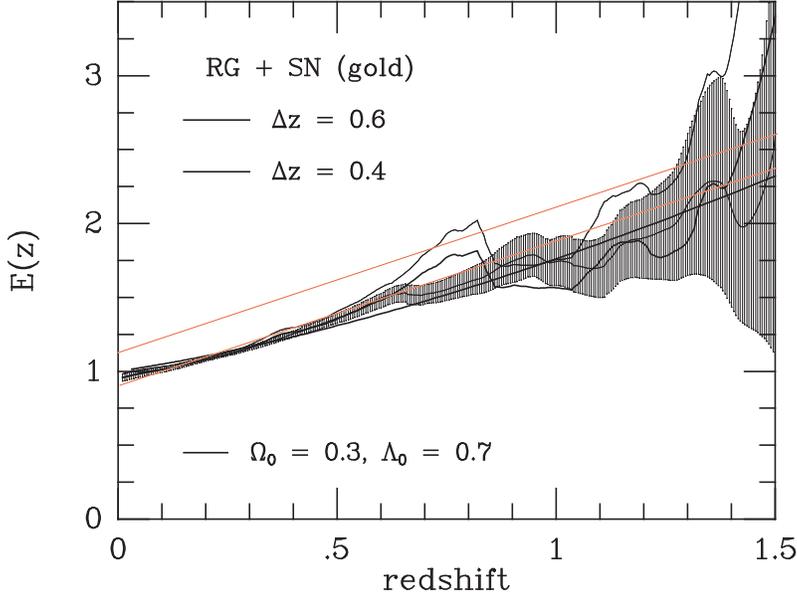}
\end{center}
\caption{The dimensionless expansion rate vs. redshift \cite{Daly
Djorgovski 2004}, $E(0) = 0.97 \pm 0.03$. An area between two red
lines corresponds to possible values of dimensionless expansion
rate in quantum model with fluctuations.} \label{fig:5}
\end{figure}

The proposed approach to the explanation of observed dimming of
some SNe Ia may provoke objections in connection with the problem
of large-scale structure formation in the universe, since the
energy density $\langle \rho_{tot} \rangle$ in the form (\ref{30})
cannot ensure an existence of a growing mode of the density
contrast $\delta \langle \rho_{tot} \rangle/\langle \rho_{tot}
\rangle$ (see e.g. \cite{Weinberg 1972}). As we have already
mentioned above the density $\langle \rho_{tot} \rangle$
(\ref{30}) describes only homogenized properties of the universe
as a whole. It cannot be used in calculations of fluctuations of
energy density about the mean value $\langle \rho_{tot} \rangle$.
Under the study of large-scale structure formation one should
proceed from the more general expression for the energy density
(\ref{29}). Defining concretely the contents of matter/energy $M$,
as for instance in the model of creation of matter and energy
proposed in \cite{Kuzmichev Kuzmichev 2004b}, one can make
calculations of density contrast as a function of redshift. The
ways to solve the problem of large-scale structure formation in
the quantum model are roughly outlined in \cite{Kuzmichev
Kuzmichev 2002}.

\subsection{Dark matter and dark energy prediction}
The quantum model allows to calculate the percentage of the
mass-energy constituents in the total energy density. For a flat
universe it predicts 29~\% for the matter density and 71~\% for
the dark energy contribution (for details see \cite{Kuzmichev
Kuzmichev 2004b}).

In Figures 6 and 7 the theoretical values of matter density and
dark energy density in comparison with observational data
summarized by \\ \cite{Spergel et al. 2003} are shown.

There is a good agreement between combined observational data and
the theoretical prediction.\\[0.5cm]

\noindent \textbf{Acknowledgements.} I would like to thank the
organizers (Bogolyubov Institute for Theoretical Physics, National
Academy of Sciences of Ukraine, Ukrainian Physical Society,
University of California), especially Professor L. Jenkovszky, for
the invitation to participate in the workshop.

\begin{figure}[ht]
\begin{center}
\includegraphics*[scale=1.4]{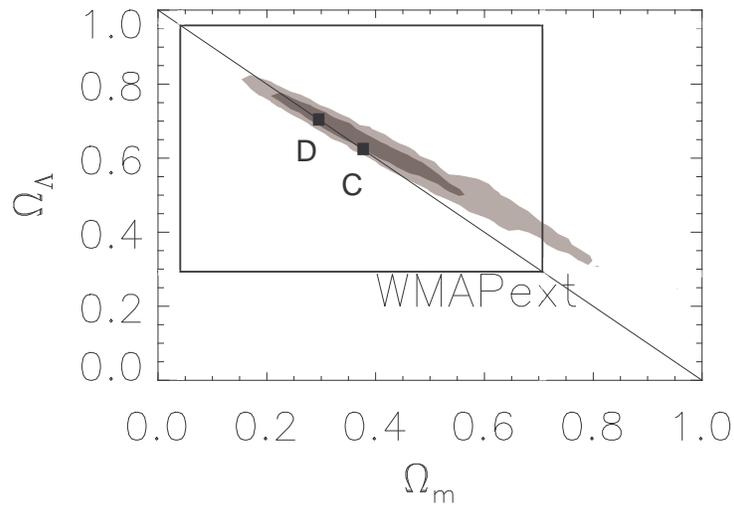}
\end{center}
\caption{The plane $\Omega_{\Lambda} \equiv
\Omega_{X}/\Omega_{tot}$ vs. $\Omega_{m} \equiv
\Omega_{M}/\Omega_{tot}$. Constraints on the density components
determined using WMAP + other CMB experiments \cite{Spergel et al.
2003}. The acceptable values of $\Omega_{\Lambda}$ and
$\Omega_{m}$ lie on the diagonal of rectangle. The central value
of the region is shown as a solid box C. The point D corresponds
to the case $Q \simeq 10$ GeV, $Q$ is the energy released in decay
of elementary quantum excitation of the vibrations of the scalar
field.} \label{fig:6}
\end{figure}

\begin{figure}[ht]
\begin{center}
\includegraphics*[scale=1.4]{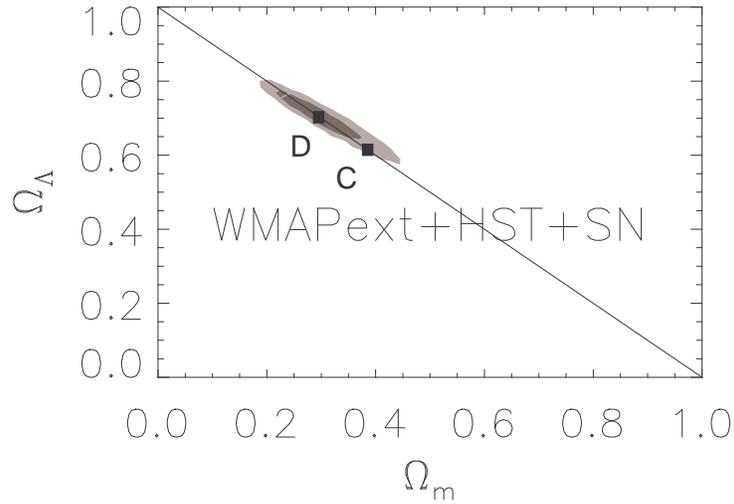}
\end{center}
\caption{Constraints on the density of matter $\Omega_{m}$ and
dark energy $\Omega_{\Lambda}$ determined using WMAPext + HST key
project data + supernova data \cite{Spergel et al. 2003}. The rest
as in Fig.~6.} \label{fig:7}
\end{figure}

\end{document}